# Evaluation of the clinical usefulness of modulated Arc treatment

**Young Kyu LEE, Hong Seok JANG, Yeon Sil KIM, Byung Ock CHOI and Young-Nam KANG[*]**

*Department of Radiation Oncology, Seoul St. Mary's Hospital, College of Medicine, The Catholic University of Korea*

**Sang Hee NAM**

*Department of Biomedical Engineering, Inje University of Korea*

**Hyeong Wook PARK**

*Department of Radiation Oncology, Yeouido St. Mary's Hospital, College of Medicine, The Catholic University of Korea*

**Shin Wook KIM, Hun Joo SHIN**

*Department of Radiation Oncology, Inchoen St. Mary's Hospital, College of Medicine, The Catholic University of Korea*

**Jae Choon LEE**

*Department of Medical Physics, Kyunggi University of Korea*

**Ji Na KIM**

*Department of Biomedical Engineering, College of Medicine, The Catholic University of Korea*

**Sung Kwang PARK**

*Department of Radiation Oncology, Busan Paik Hospital, Inje University*

**Jin Young KIM**

*Department of Radiation Oncology, Haeundae Paik Hospital, Inje University*

abstract
The purpose of this study is to evaluate the clinical usefulness of modulated arc (mARC) treatment techniques. The mARC treatment plans of the non-small cell lung cancer (NSCLC) patients were performed in order to verify the clinical usefulness of mARC. A pre study was conducted to find the most competent plan condition of mARC treatment and the usefulness of mARC treatment plan was evaluated by comparing it with the other Arc treatment plans such as Tomotherapy and RapidArc. In the case of mARC, the optimal condition for the mARC plan was determined by comparing the dosimetric performance of the mARC plans with the use of various parameters. The various parameters includes the photon energies (6 MV, 10 MV), optimization point angle (6°-10° intervals), and total segment number (36-59 segment). The best dosimetric performance of mARC was observed at 10 MV photon energy and the point angle 6 degree, and 59 segments. The each treatment plans of three different techniques were compared with the following parameters: conformity index (CI), homogeneity index (HI), target coverage, dose in the OARs, monitor units (MU), beam on time and the normal tissue complication probability (NTCP). As a result, all three different treatment techniques show the similar target coverage. The mARC results the lowest $V_{20}$ and MU per fraction compared with both RapidArc and Tomotherapy plan. The mARC plan reduces the beam on time as well. Therefore, the results of this study provided a satisfactory result which mARC technique is considered as a useful clinical technique for radiation treatment.

Keywords: modulated Arc treatment, RapidArc, Tomotherapy , Non-small cell lung cancer

Email: ynkang33@gmail.com
Fax: +82-2-2258-2532

## I. INTRODUCTION

Helical Tomotherapy (Accuray Inc., Madison, WI) is an arc-based, photon intensity modulated radiotherapy (IMRT) delivery system that utilizes a rotating helical fan beam with a multi-leaf collimator (MLC) containing 64 binary leaf-pairs [1]. RapidArc (Varian Medical Systems, Palo Alto, CA) is also an arc-based photon IMRT delivery system, but the entire treatment volume can be covered by a single rotation of the gantry unlike helical tomotherapy. As an alternative to serial tomotherapy, IMAT utilized multiple coplanar arcs with constant gantry speed, constant dose rate, continuous gantry rotation and dynamic MLC, but each arc delivered only one level of intensity [2]. RapidArc is capable of varying the MLC leaf position, dose rate and gantry rotation speed simultaneously during delivery. These techniques achieved significantly better target dose homogeneity while maintaining acceptable sparing of organs at risk (OARs) [3].

The mARC (Prowess Panther & Siemens, Concord, CA) technique has recently become available for Siemens ARTISTE linear accelerators as an analogue to RapidArc and VMAT (Elekta Oncology Systems, Crawley, West Sussex, UK) available for Varian and Elekta machines [4-6]. All three techniques offer highly conformal treatment, since those techniques have a large number of beam directions, which may create a complete or partial gantry rotation. Moreover the beam on time decreases notably due to continuous gantry rotation, MLC movement and small segment number [7]. The purpose of this study therefore is to evaluate the clinical usefulness of mARC treatment techniques for NSCLC case. A pre study was conducted to find the most competent plan condition of mARC treatment and the usefulness of mARC treatment plan was evaluated by comparing it with the other Arc treatment plans such as Tomotherapy and RapidArc.

## II. Material and method

Three types of radiation treatment plans, mARC, Tomotherapy and RapidArc plans, were performed for 10 NSCLC patients. Chest CT images were obtained with 3 mm slice thickness. Gross tumor volume (GTV) and planning target volume (PTV) were delineated by one qualified radiation oncologist. PTV1 is generated from a constant 1.5 cm cranio-caudal margin and a constant 1 cm radial margin of GTV1 (supraclavicular mass). PTV2 is generated from a constant 1.5 cm cranio-caudal margin and a constant 1 cm radial margin of GTV2 (mediastinum). The prescribed dose to the PTV1 and PTV2 was 6000 cGy in 30 fractions over 6 weeks. Prowess Panther™ 5.2 (Prowess Panther & Siemens, Concord, CA) for mARC, HiART™ Tomotherapy Planning Station 4.2.3 (Accuray Inc., Madison, WI) for Tomotherapy, and Eclipse™ Treatment Planning System 10.0 (Varian Medical Systems, Palo Alto, CA) for RapidArc were used as a treatment planning system.

### 2.1 mARC

Conventional arc treatment modalities such as RapidArc and VMAT, those have continuous dose delivery function. In other words, they can deliver radiation dose continuously while the gantry and MLC are in motion. mARC, by contrast, only implement radiation beam on while the MLC remains static, although the gantry moves along an arclet of a few degrees. The mARC beam is switched on in a small arclet angle around each optimization point angle [7]. Figure 1 shows schematic illustration of mARC. In the case of mARC, the optimal condition for the mARC plan was determined by comparing the dosimetric performance of the mARC plans with the use of various parameters. The various parameters includes the photon energies (6 MV, 10 MV), optimization point angle (6°-10° intervals), and total segment number (36-59 segment). The optimized mARC plan is compared with the conventional arc therapy techniques such as Helical Tomotherapy and RapidArc.

2.2 Conventional arc therapy

In Tomotherapy plan case, main parameters such as the field width, pitch, and modulation factor were fixed, i.e. 1-cm field width, a 0.3 pitch, and a 2 modulation factor. The isocenter was placed at the center of summation structure of PTV1 and PTV2. For the RapidArc, Two arcs treatment plans with the same isocenter were performed. In the case of tumor located on the left side, the first arc rotated counterclockwise from gantry angle 179°–359° and the second arc rotated clockwise from 359° to 179°. A mirrored configuration was used when the tumor is located on the right side. The total arc angle was 718° degree. The two arc technique was expected to achieve better target dose coverage than the single arc because the independent optimization of two arcs allows each arc create a completely unrelated sequence of MLC shapes, dose rates and gantry speed combinations. It is recommended when the target and OARs are closely located so that the treatment planning has some limitations.

2.3 Comparison of the three different techniques

The each treatment plans of three different techniques were compared with the following parameters: conformity index (CI), homogeneity index (HI), target coverage, dose in the OARs, monitor units (MU), beam on time and the normal tissue complication probability (NTCP). The conformity index (CI) was defined as:

$$\text{CI} = \frac{V_{PTV}}{TV_{PV}} \times \frac{V_{TV}}{TV_{PV}} \qquad (1)$$

Where $V_{TV}$ is the treatment volume of the prescribed isodose lines and $V_{PTV}$ is the volume of the PTV. $TV_{PV}$ is the volume of $V_{PTV}$ within the $V_{TV}$. It represents that the smaller the CI value, the better conformal the PTV fitting [8]. The homogeneity index (HI) was defined as:

$$\text{HI} = \frac{(D_{2\%} - D_{98\%})}{D_{median}} \qquad (2)$$

Homogeneity Index (HI) is defined as the maximum dose delivered to 2% of the target volume ($D_{2\%}$) minus the dose delivered to 98% of the target volume ($D_{98\%}$) divided by the median dose ($D_{median}$) to the target volume. The average cumulative DVH was generated for the PTV and OARs from the individual DVHs. Paired two tailed t-test was used to detect any statistically significant differences in these parameters among the 3 different techniques [9]. The NTCP was used the Lyman NTCP model with parameters described by Burman et al. and Lyman [10, 11].

## III. Result and discussion

3.1 Target coverage

The mean volume of the PTV1 and PTV2 was 42.3 ± 3.162 cc and 51.3 ± 11.597 cc. In the case of mARC, the best dosimetric performance of mARC was observed with the main factors of 10 MV photon energy and the point angle 6 degree, 59 segments. The values of the study among the various parameters in mARC were tabulated in Table 1. The dose color wash from 24 Gy (40%) to 66 Gy (110%) of the three arc techniques were illustrated in Figure 2. Both mARC and tomotherapy plans had a better conformity compared with RapidArc, with a relative improvement of 28% and 29% in CI of PTV1, and with a relative improvement of 32% and 7% in CI of PTV2. However, RapidArc produced the best dose homogeneity, as reflected in the least formula 1, and it was lower than that of mARC and tomotherapy plan (Table 2).

3.2 Organs at risks

mARC produced the lowest V20 (volume of lung-PTV receiving > 20 Gy). The V20 for mARC was 6%, with an absolute difference of 1.3% lower than tomotherapy and 0.2% lower than RapidArc. Mean lung dose was 6.45 Gy, 0.16 Gy lower than tomotherapy. However, 0.2 Gy higher than RapidArc. The mean esophagus dose was 4.07 Gy, with an absolute difference of 4.15 Gy lower than RapidArc, it was similar in mARC and tomotherapy plan. The mean airway dose of tomotherapy plan was 14.87 Gy, which was the lowest among the three techniques (4.43 Gy lower than mARC and 4.41 Gy lower than RapidArc). The dose volume histograms for the PTV and OAR with different techniques were shown in Figures. 3 and 4.

3.3 Monitor units (MU) and beam-on time

The mARC results the lowest V20 and MU per fraction compared with both RapidArc and Tomotherapy plan as decreasing 214 MU and 81 MU, respectively.

The treatment time using mARC technique was 307.4s, and was shorter than tomotherapy plan by 498s. RapidArc plan gave the shortest treatment time, with a treatment time of 188.5s.

## Ⅳ. Conclusion

The target coverages were similar in all three different treatment techniques. Among those, the RapidArc produced the best dose homogeneity index. However, the conformity index of RapidArc were 1.510 at PTV1 and 1.632 at PTV2, of Tomotherapy were 1.072 at PTV1 and 1.520 at PTV2, of mARC were 1.100 at PTV1 and 1.116 at PTV2. The mARC produced the lowest V20. However, RapidArc produced the best mean lung dose. Both mARC and RapidArc plans had a shorter treatment time, compared with tomotherapy, relatively fast of 62% and 77% in treatment time. Therefore, the results of this study provided a satisfactory result which mARC technique is considered as a useful clinical technique for radiation treatment.

## ACKNOWLEDGEMENT

We would like to acknowledge the financial support from the R&D Convergence Program of MSIP (Ministry of Science, ICT and Future Planning) and ISTK (Korea Research Council for Industrial Science and Technology) of Republic of Korea (B551179-12-08-00).

# REREFENCES

Table 1. Summary of the study parameters the mARC treatment technique

| | | Point angle | segment | Lung V20 | Airway Mean | Cord Max | Thyroid Mean | Esophagus Mean | Heart Mean | PTV1 D90 | PTV2 D90 |
|---|---|---|---|---|---|---|---|---|---|---|---|
| | | | | % | Gy | | | | | % | |
| mARC | 6MV | 6 | 59 | 6.0 | 20.402 | 19.668 | 9.015 | 4.579 | 1.398 | 96.7 | 95.1 |
| | | 7 | 51 | 6.0 | 21.694 | 18.061 | 10.344 | 6.923 | 1.287 | 95.2 | 95.1 |
| | | 8 | 45 | 6.1 | 21.784 | 20.694 | 11.824 | 6.078 | 1.315 | 96.3 | 94.3 |
| | | 9 | 40 | 6.9 | 24.004 | 22.645 | 15.145 | 8.631 | 1.348 | 94.4 | 94.2 |
| | | 10 | 36 | 7.7 | 24.477 | 20.681 | 18.769 | 8.462 | 1.389 | 94.1 | 93.4 |
| | 10MV | 6 | 59 | 6.0 | 19.503 | 19.602 | 8.856 | 4.071 | 1.201 | 97.8 | 94.6 |
| | | 7 | 51 | 6.1 | 21.600 | 20.272 | 9.518 | 6.043 | 1.131 | 95.9 | 94.8 |
| | | 8 | 45 | 6.1 | 21.254 | 19.835 | 10.334 | 6.455 | 1.234 | 96.8 | 95.4 |
| | | 9 | 40 | 7.0 | 23.727 | 25.687 | 13.641 | 8.042 | 1.115 | 95.3 | 95.3 |
| | | 10 | 36 | 6.5 | 25.090 | 21.844 | 16.991 | 8.253 | 1.182 | 95.2 | 93.9 |

Table 2. Summary of the study parameters among the 3 different treatment techniques

|  | mARC | RapidArc | Tomotherapy | P value |
|---|---|---|---|---|
| Conformity index | PTV1: 1.100± 0.271 | PTV1: 1.510± 0.042 | PTV1: 1.072± 0.061 | a:p <0.756<br>b:p <0.001 |
|  | PTV2: 1.116± 0.152 | PTV2: 1.632± 0.047 | PTV2: 1.520± 0.110 | a:p <0.001<br>b:p <0.001 |
| Homogeneity index | PTV1: 0.123± 0.021 | PTV1: 0.06± 0.027 | PTV1: 0.228± 0.002 | a:p <0.001<br>b:p <0.013 |
|  | PTV2: 0.092± 0.011 | PTV2: 0.098± 0.003 | PTV2: 0.030± 0.008 | a:p <0.001<br>b:p <0.174 |
| Mean lung-PTV dose (Gy) | 6.45± 0.229 | 6.25± 0.177 | 6.61± 0.146 | a:p <0.081<br>b:p <0.038 |
| V20 (lung-PTV) (%) | 6.0± 0.462 | 6.2± 0.255 | 7.3± 0.970 | a:p <0.002<br>b:p <0.383 |
| V10 (lung-PTV) (%) | 26.0± 0.316 | 23.3± 0.158 | 24.8± 0.101 | a:p <0.001<br>b:p <0.001 |
| V5 (lung-PTV) (%) | 41.0± 0.792 | 38.6± 0.158 | 39.4± 1.029 | a:p <0.041<br>b:p <0.001 |
| Mean (esophagus) (Gy) | 4.07± 0.071 | 8.22± 0.151 | 4.05± 0.079 | a:p <0.739<br>b:p <0.001 |
| Mean (Airway) (Gy) | 19.3± 0.255 | 19.5± 0.412 | 14.87± 0.016 | a:p <0.001<br>b:p <0.399 |
| Treatment time (s) | 307.4± 7.515 | 188.5± 8.214 | 805.4± 40.343 | a:p <0.001<br>b:p <0.001 |
| Monitor units | 465.9± 20.210 | 680.3± 15.951 | 547.4± 30.592 | a:p <0.001<br>b:p <0.001 |
| NTCP (lung-PTV) (%) | 7.68± 0.608 | 7.45± 0.707 | 7.85± 0.453 | a:p <0.074<br>b:p <0.083 |

Abbreviations: a = mARC vs. Tomotherapy, b = mARC vs. RapidArc

Figure Captions.

Figure 1. Schematic illustration of mARC

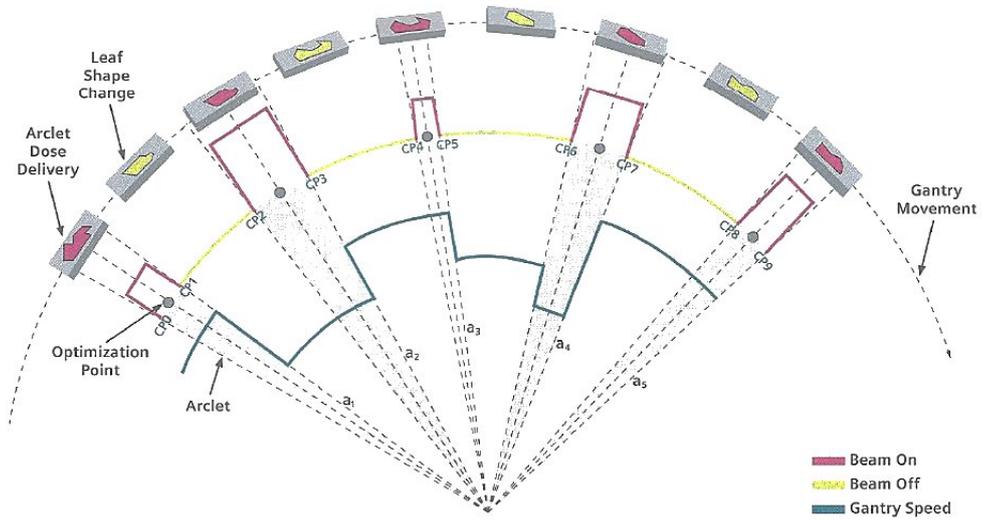

Figure 2. Dose distribution of the three different treatment techniques in a representative case.

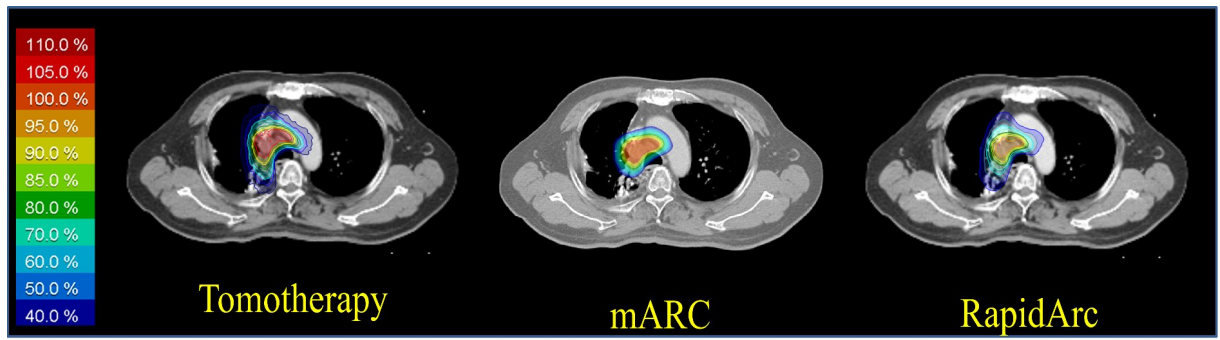

Figure 3. The dose volume histogram for planning target volume of the three different treatment techniques.

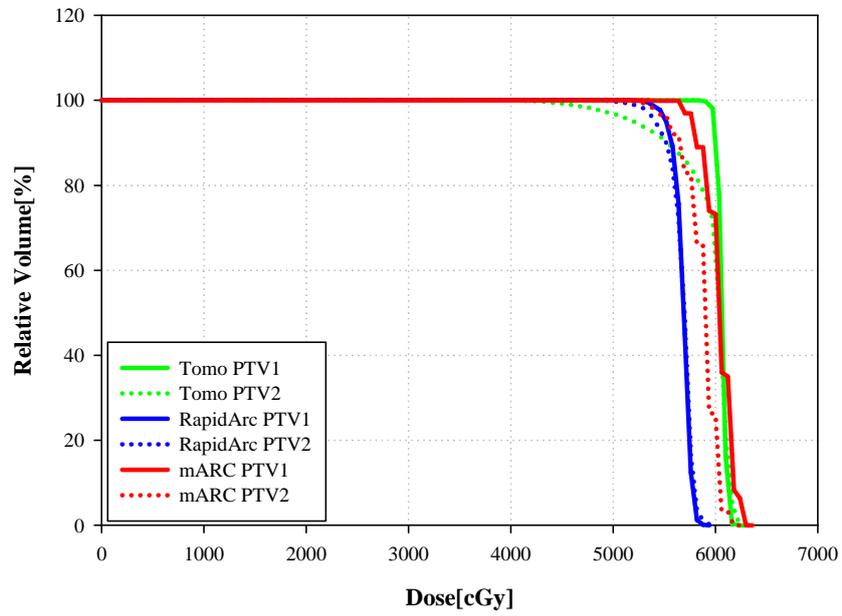

Figure 4. The dose volume histogram for normal organ (lung, esophagus, airway) of the three different treatment techniques.

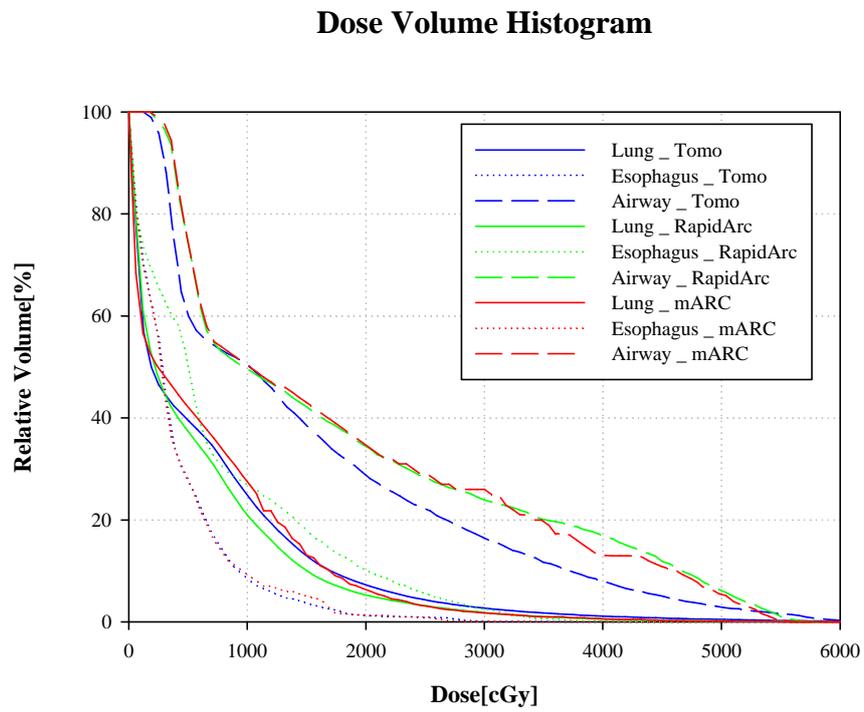

Table 1. Summary of the study parameters the mARC treatment technique

| | | Point angle | segment | Lung V20 | Airway Mean | Cord Max | Thyroid Mean | Esophagus Mean | Heart Mean | PTV1 D90 | PTV2 D90 |
|---|---|---|---|---|---|---|---|---|---|---|---|
| | | | | % | Gy | | | | | % | |
| mARC | 6MV | 6 | 59 | 6.0 | 20.402 | 19.668 | 9.015 | 4.579 | 1.398 | 96.7 | 95.1 |
| | | 7 | 51 | 6.0 | 21.694 | 18.061 | 10.344 | 6.923 | 1.287 | 95.2 | 95.1 |
| | | 8 | 45 | 6.1 | 21.784 | 20.694 | 11.824 | 6.078 | 1.315 | 96.3 | 94.3 |
| | | 9 | 40 | 6.9 | 24.004 | 22.645 | 15.145 | 8.631 | 1.348 | 94.4 | 94.2 |
| | | 10 | 36 | 7.7 | 24.477 | 20.681 | 18.769 | 8.462 | 1.389 | 94.1 | 93.4 |
| | 10MV | 6 | 59 | 6.0 | 19.503 | 19.602 | 8.856 | 4.071 | 1.201 | 97.8 | 94.6 |
| | | 7 | 51 | 6.1 | 21.600 | 20.272 | 9.518 | 6.043 | 1.131 | 95.9 | 94.8 |
| | | 8 | 45 | 6.1 | 21.254 | 19.835 | 10.334 | 6.455 | 1.234 | 96.8 | 95.4 |
| | | 9 | 40 | 7.0 | 23.727 | 25.687 | 13.641 | 8.042 | 1.115 | 95.3 | 95.3 |
| | | 10 | 36 | 6.5 | 25.090 | 21.844 | 16.991 | 8.253 | 1.182 | 95.2 | 93.9 |

Table 2. Summary of the study parameters among the 3 different treatment techniques

|  | mARC | RapidArc | Tomotherapy | P value |
|---|---|---|---|---|
| Conformity index | PTV1: 1.100± 0.271 | PTV1: 1.510± 0.042 | PTV1: 1.072± 0.061 | a:p <0.756<br>b:p <0.001 |
|  | PTV2: 1.116± 0.152 | PTV2: 1.632± 0.047 | PTV2: 1.520± 0.110 | a:p <0.001<br>b:p <0.001 |
| Homogeneity index | PTV1: 0.123± 0.021 | PTV1: 0.06± 0.027 | PTV1: 0.228± 0.002 | a:p <0.001<br>b:p <0.013 |
|  | PTV2: 0.092± 0.011 | PTV2: 0.098± 0.003 | PTV2: 0.030± 0.008 | a:p <0.001<br>b:p <0.174 |
| Mean lung-PTV dose (Gy) | 6.45± 0.229 | 6.25± 0.177 | 6.61± 0.146 | a:p <0.081<br>b:p <0.038 |
| V20 (lung-PTV) (%) | 6.0± 0.462 | 6.2± 0.255 | 7.3± 0.970 | a:p <0.002<br>b:p <0.383 |
| V10 (lung-PTV) (%) | 26.0± 0.316 | 23.3± 0.158 | 24.8± 0.101 | a:p <0.001<br>b:p <0.001 |
| V5 (lung-PTV) (%) | 41.0± 0.792 | 38.6± 0.158 | 39.4± 1.029 | a:p <0.041<br>b:p <0.001 |
| Mean (esophagus) (Gy) | 4.07± 0.071 | 8.22± 0.151 | 4.05± 0.079 | a:p <0.739<br>b:p <0.001 |
| Mean (Airway) (Gy) | 19.3± 0.255 | 19.5± 0.412 | 14.87± 0.016 | a:p <0.001<br>b:p <0.399 |
| Treatment time (s) | 307.4± 7.515 | 188.5± 8.214 | 805.4± 40.343 | a:p <0.001<br>b:p <0.001 |
| Monitor units | 465.9± 20.210 | 680.3± 15.951 | 547.4± 30.592 | a:p <0.001<br>b:p <0.001 |
| NTCP (lung-PTV) (%) | 7.68± 0.608 | 7.45± 0.707 | 7.85± 0.453 | a:p <0.074<br>b:p <0.083 |

Abbreviations: a = mARC vs. Tomotherapy, b = mARC vs. RapidArc

Figure Captions.

Figure 1. Schematic illustration of mARC

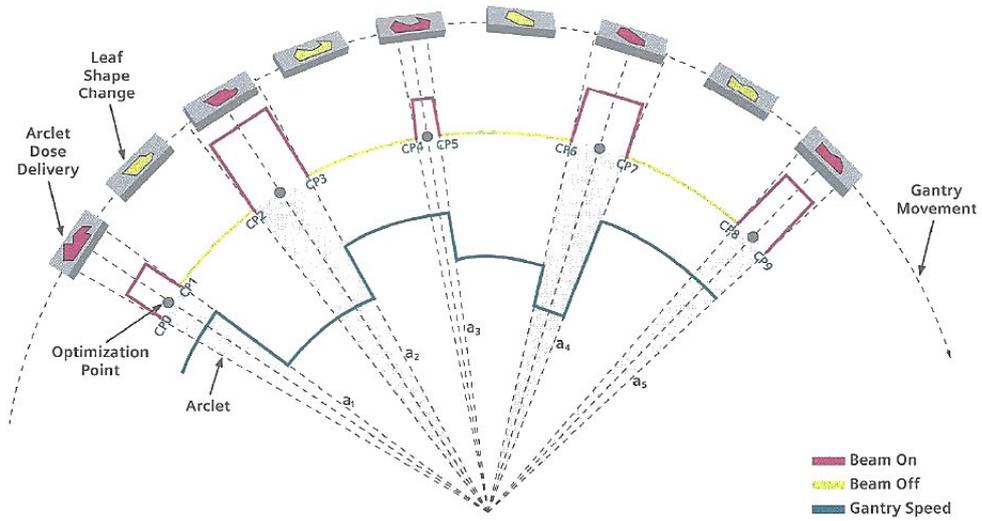

Figure 2. Dose distribution of the three different treatment techniques in a representative case.

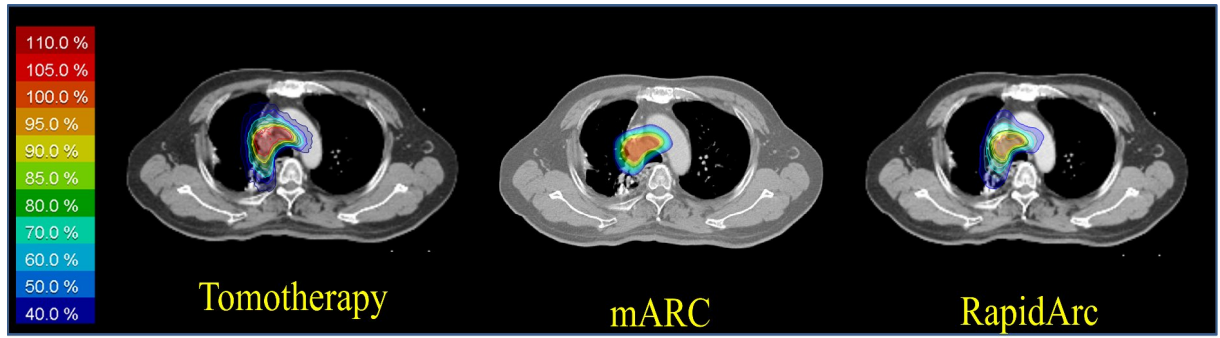

Figure 3. The dose volume histogram for planning target volume of the three different treatment techniques.

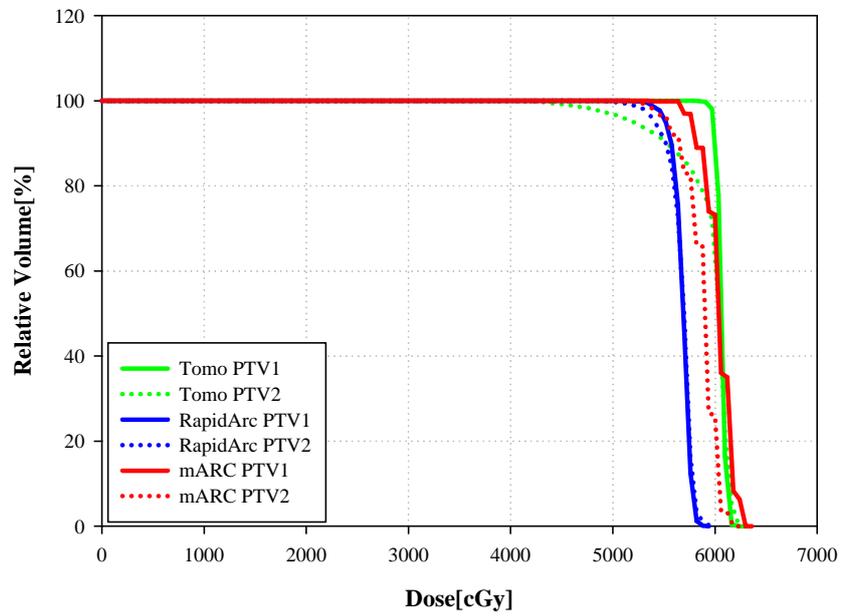

Figure 4. The dose volume histogram for normal organ (lung, esophagus, airway) of the three different treatment techniques.

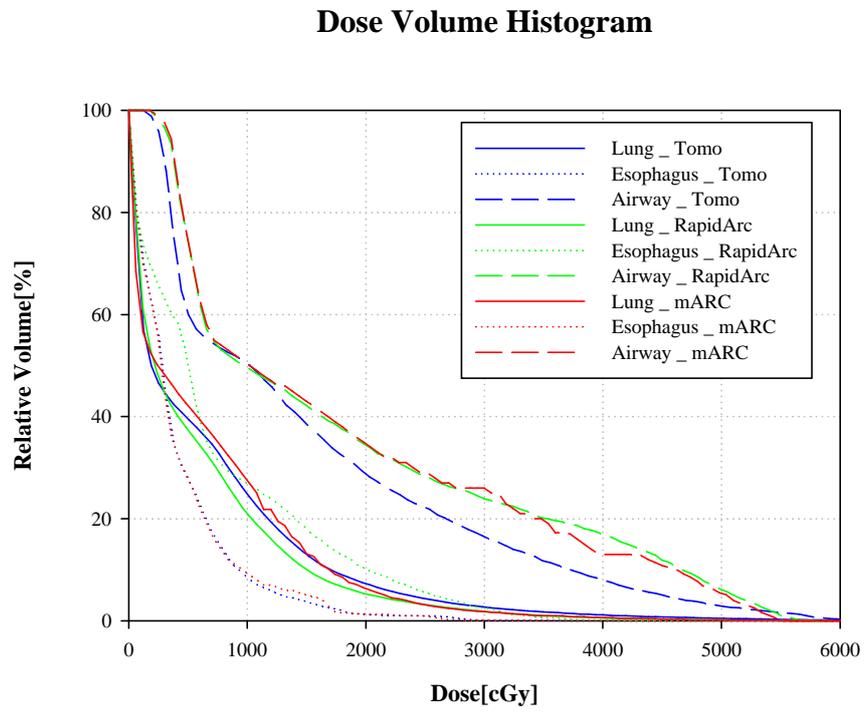